\documentclass[prb,aps,twocolumn,showpacs,amsmath,amssymb]{revtex4}

\usepackage{epsf}

\begin{document}
\draft
\newcommand{\ve}[1]{\boldsymbol{#1}}

\title{Magnetism and superconductivity at LAO/STO-interfaces: \\
the role of Ti $3d$ interface electrons}

\author{N.~Pavlenko$^{1,2}$, T.~Kopp$^2$, E.Y.~Tsymbal$^3$, G.A.~Sawatzky$^4$, and J.~Mannhart$^5$}
\address{$^1$Institute for Condensed Matter Physics, 79011 Lviv, Ukraine,\\
$^2$ EKM and Institut f\"ur Physik, Universit\"at Augsburg, 86135 Augsburg, Germany \\
$^3$ Department of Physics and Astronomy, Nebraska Center for Materials and Nanoscience, University of 
Nebraska, Lincoln, Nebraska 68588-0299, USA \\
$^4$Department of Physics and Astronomy, University of British Columbia, Vancouver, Canada V6T1Z1
$^5$Max Planck Institute for Solid State Research, 70569 Stuttgart, Germany 
}

\begin{abstract}
Ferromagnetism and superconductivity are in most cases adverse. However, recent experiments reveal that they 
coexist at interfaces of LaAlO$_3$ and SrTiO$_3$. We analyze the magnetic state within density functional 
theory and provide evidence that magnetism is not an intrinsic property of the two-dimensional electron 
liquid at the interface. We demonstrate that the robust ferromagnetic state
is induced by the oxygen vacancies in SrTiO$_3$- or in the LaAlO$_3$-layer. This 
allows for the notion that areas with increased density of oxygen vacancies produce ferromagnetic puddles and 
account for the previous observation of a superparamagnetic behavior in the superconducting state.
\end{abstract}


\date{\today}

\maketitle

The formation of a metallic state at the interface of the bulk insulators LaAlO$_3$ and SrTiO$_3$~\cite{hwang} 
has become a prototype for the reconstruction of electronic states in systems with artificially reduced 
dimensionality. This two-dimensional (2D) electronic system is affected by sizable electronic correlations 
which allow characterizing the extended interface electronic states as an electron 
liquid~\cite{breitschaft,eom}. The correlations not only induce a superconducting state~\cite{reyren} but also 
support magnetism~\cite{brinkman}. A novel yet unexplained phenomenon is the coexistence of magnetism and 
superconductivity \cite{dikin,li} in the 2D electron liquid. 

Recent measurements by Dikin {\it et al.}~\cite{dikin} demonstrate a hysteretic behavior in the 
field-dependences of magnetoresistance and critical temperature which suggests the existence of ferromagnetism 
in the superconducting samples of LaAlO$_3$ (LAO) grown on SrTiO$_3$ (STO). Moreover, Dikin {\it et al.}\  
take the viewpoint that two distinct electronic systems are associated with the antagonistic superconducting 
and  ferromagnetic properties: the electrons that are generated by the polar catastrophe mechanism are 
suggested to be related with magnetism while superconductivity is associated with the charge carriers  
induced by the presence of oxygen vacancies. The concept of two different types of charge carriers which could 
contribute to the interface electrical transport has been discussed also in \cite{seo,fix,popovic,ariando}. 

Very recently, Li {\it et al.}~\cite{li} probed magnetism through torque magnetometry which allows detecting directly the 
magnetic moment of the interface in an external magnetic field $H$. They  found a strong superparamagnetic 
torque signal in the superconducting state. With the assumption that the signal originates from the STO layer 
next to the interface, they obtained the magnetic moment $M$ of $0.3-0.4 \,\mu_B$ per unit 
cell and a collective magnetic moment of the superparamagnetic grains of the order of 1000~$\mu_B$. The observation 
of a superparamagnetic $M(H)$ indicates that ferromagnetic grains form even in the superconducting state. 
Magnetic oxygen sites at the AlO$_2$-surface (cf.~Ref.~\cite{elfimov}) and the build-up of triplet coupling of 
Ti $3d$ states through the oxygen bonds (or possibily vacancies) in the TiO$_2$ interface plane 
(cf.~Ref.~\cite{lau}) have been proposed~\cite{li} as scenarios for the formation of a magnetic state.

The interpretation of the experimental results opens a compelling question: can the Ti $3d$ orbitals that were
identified in the previous band structure calculations (see, e.g., 
Refs.~\cite{pentcheva2,pavlenko,popovic2,okamoto,lee,tsymbal2,dagotto,tsymbal,zhong}) be responsible both for the 
metallic and magnetic states coexisting at the same interface?
In this paper, we present the results of density functional studies which support the existence of a robust 
ferromagnetic state at the LAO/STO interface induced by oxygen vacancies. 
We demonstrate that both the magnetism and conductivity
occur involving the Ti $3d$ electrons but the magnetism is due to rather confined electrons around O vacancies while the 
conductivity is a result of the 2D electron gas caused by electronic reconstruction. We argue that this behavior is a 
prerequisite of co-existence of magnetism and superconductivity which are observed at low temperatures.

\begin{figure}[tbp]
\epsfxsize=8.5cm {\epsfclipon\epsffile{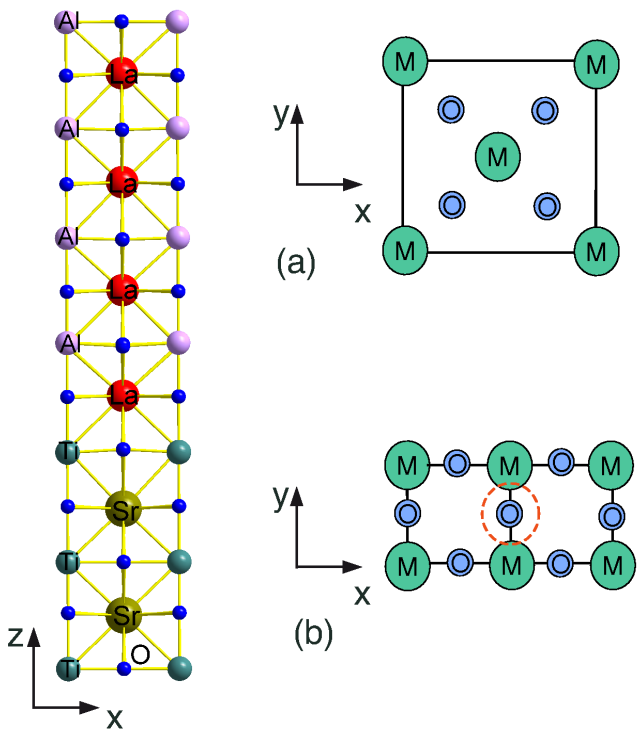}}
\caption{Schematic view of the SrTiO$_3$/LaAlO$_3$ heterostructure.
The supercell contains a 4 unit cell thick LaAlO$_3$ layer deposited on a 2.5 unit cell thick SrTiO$_3$ slab. 
The full supercell consists of two symmetric parts of the depicted structure and a vacuum layer of 13~\AA.
The structures on the right side show (a) a projection of the supercell
of STO/LAO on the $(x,y)$-plane of TiO$_2$, and (b) a M$_2$O$_4$ (M=Ti, Al)-plaquette
generated for the study of the system with O-vacancies. The position of an O-vacancy is identified by a 
red dashed circle.
}
\label{fig1}
\end{figure}

To explore whether a ferromagnetic state is induced 
at these interfaces, we consider oxygen vacancies as a mechanism responsible for magnetism. 
We generate a number of supercells which consist of two symmetric LAO/STO
parts, where each part contains a stack of 4 unit-cell (uc) thick LAO layers deposited on a STO-slab 
of a thickness varying between 1.5 uc and 6 uc. The interfacial configuration is considered as TiO$_2$/LaO. 
The LAO-STO-LAO parts are separated by a 13~\AA\, thick vacuum sheet. 
Oxygen vacancies are assumed to lie in the first interfacial TiO$_2$-layers
or in one of the AlO$_2$-layers of the LAO-film. A cell with an 
oxygen vacancy in MO$_2$ (M=Ti. Al) is sketched in 
Fig.~\ref{fig1}(b). The vacancy is introduced by excluding the oxygen atom 
O($a/2$, $b/2$) in the center of the M$_2$O$_4$-plaquette. The location of the vacancy at the interface is 
motivated by the experimental evidence of O-vacancies present in STO in samples grown at oxygen pressures 
below 10$^{-5}$~mbar \cite{herranz,kalabukhov}. 

Density functional calculations are performed using the Generalized Gradient Approximation (GGA) in the 
Perdew-Burke-Ernzerhof pseudopotential implementation \cite{pbe} in the Quantum Espresso (QE) 
package \cite{qe}. We use a kinetic energy cutoff of 640~eV and the Brillouin zone
of the 106- to 166-atom supercells sampled with 5$\times$5$\times$1 to 9$\times$9$\times$1 $\ve{k}$-point 
grids. An increase of the k-point mesh from (5x5x1) to (7x7x1) leads
to a negligibly small change of the total energy by 0.005Ry and to an increase
of the Ti magnetic moments by small values of about 0.05~$\mu_B$ in the presence of O-vacancies. 
The difference between the Ti magnetic moments for the two different (2nx2nx1) 
and (nx2nx1) k-meshes for n=4 is about 0.02 $\mu_B$, which is a negligibly small value.
In our calculations we account for a local Coulomb repulsion of Ti $3d$ electrons
by employing a GGA+$U$ approach with $U_{\rm Ti}=2$~eV~\cite{breitschaft}.
First, we consider pure stoichiometric TiO$_2$/LaO-interfaces as reference
for the oxygen doped interfaces. The supercells 
which contain (2$\times$2) planar unit cells have been structurally relaxed along the $z$-axis by a 
combination of the  optimization procedures of the full potential WIEN2k-package and the pseudopotential QE 
package \cite{wien2k,qe}. The in-plane lattice constants have been fixed to their bulk-STO cubic values ($a_{{\rm 
STO}}=b_{{\rm STO}}=3.905~$\AA). Similar to the previous calculations
\cite{pentcheva2,pavlenko,popovic2,okamoto,lee,tsymbal2}, we find that a metallic state 
is produced at the LAO/STO interface due to the electronic reconstruction. 

\begin{figure}[tbp]
\epsfxsize=8.5cm {\epsfclipon\epsffile{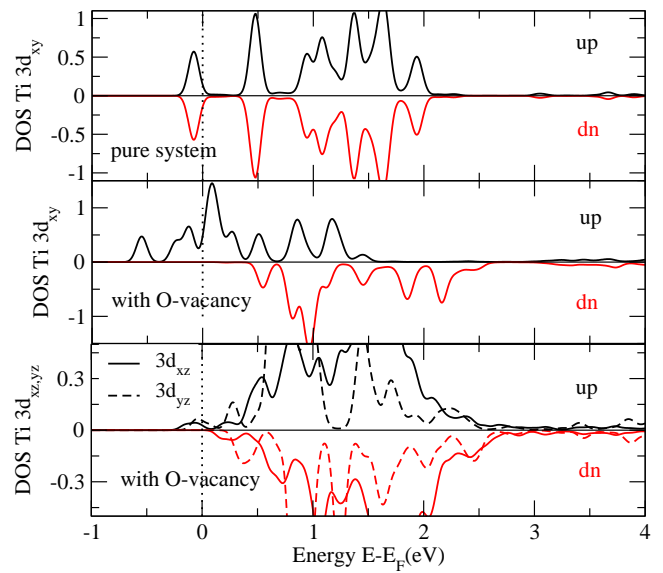}}
\caption{Projected DOS (in eV$^{-1}$) for $3d$ ($t_{2g}$) states of the interfacial Ti
in the supercell containing 4 unit cell-thick LaAlO$_3$ layers and a 4-unit cells thick SrTiO$_3$ layer.
DOS in the pure system and in the system with one O-vacancy (25\%) per supercell area in  the interfacial
TiO$_2$-layer are shown for comparison. The vertical grey line denotes the Fermi level.
} 
\label{fig2}
\end{figure}

Fig.~\ref{fig2} presents the projected Ti $3d$ densities of states (DOS) for both spin directions in the system with 
supercells
containing $4$~LAO uc and 4 STO uc along the $z$-direction (the full supercell contains twice the number of
LAO unit cells). The difference in the spin-projected DOS implies a non-zero spin 
polarization. For the pure system without oxygen vacancies, the occupancies of the spin-up and spin-down 
$3d$ states are almost identical. 
The maximal magnetic moments of the interface Ti are about $0.005\,\mu_B$ and the polarization
from the more distant TiO$_2$ planes is negligible. 
The calculated magnetic moment
per (1$\times$1) unit cell of LAO/STO interface is 0.03~$\mu_B$ which originates mostly from the
surface oxygen sites. 
This polarization is too small to support a robust ferromagnetic state suggesting that magnetism is not due to the
pure interface electron gas. 

\begin{figure}[tp]
\epsfxsize=8.5cm {\epsfclipon\epsffile{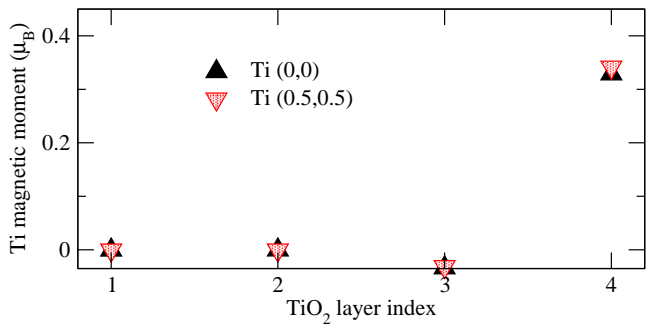}}
\caption{Magnetic moments of Ti atoms in 
different TiO$_2$ layers in the (4~uc)SrTiO$_3$/(4~uc)LaAlO$_3$ structures
for the structure with one O-vacancy in the interfacial TiO$_2$ layer.
The black up and red down triangles  correspond to the two
Ti atoms with the planar coordinates (0,0) and ($0.5a$, $0.5a$) in a doubled unit cell of SrTiO$_3$.
The TiO$_2$ layer 4 is the layer next to the interface.
} \label{fig3}
\end{figure}

The situation with O-vacancies is different.
An oxygen vacancy adds two extra electrons at the interface to preserve charge neutrality. 
The two electrons are most
likely localized in the vicinity of the O-vacancy as found for CaO by Elfimov {\it et al.}~\cite{elfimov}.
As shown below, this enhances the charge density and increases the exchange 
splitting of the spin bands; consequently O-vacancies stabilize the ferromagnetic order. 

First, we assume that the oxygen vacancy lies within the TiO$_2$ plane at the interface. 
In the oxygen-deficient system, we find sizable Ti magnetic moments at the interfacial TiO$_2$-plane (see 
Fig.~\ref{fig3}). The 
magnetic moment of the Ti atoms next to the O-vacancy is $\sim 0.33\,\mu_B$ and that of the more distant Ti(0,0) 
atoms at the interfacial plane is $\sim0.34\,\mu_B$. Magnetic moments on Ti atoms away 
from the interface are negligible (Fig.~\ref{fig3}). We also find a 
sizable magnetic moment on the AlO2 surface plane of about $-0.18\,\mu_B$ 
aligned antiparallel to the magnetic moment of the interface Ti atoms. 
Needless to say, the concentration 
of O-vacancies in these model structures is far higher than the average density in the experimentally investigated
heterostructures. Nevertheless, it is evident that a triplet coupling is induced between the nearest-neighbor 
Ti sites and that ferromagnetism is enhanced in O-vacancy rich regions of the interfacial plane. 

The elimination of the central oxygen in the ($2 \times 1$) configuration results in the formation of
stripes of O-vacancies along the $y$-direction characterized by two vacancies near Ti($0.5a$, $0$) and no vacancies
near Ti(0,0) atom.
To test the stability of such an ``inhomogeneous'' distribution of O-vacancies, we 
have also performed GGA+$U$ calculations of a supercell with an ordered `homogeneous" 
arrangement of the vacancies corresponding to exactly one vacancy per each Ti atom. 
This can be obtained by elimination of one oxygen in the 
square ($\sqrt{2}\times \sqrt{2}$)-supercell shown in Fig.~\ref{fig1}(a).  
The comparison of the calculated total energy with the 
energy for that of the  ($2 \times 1$)-supercell (Fig.~\ref{fig1}(b)), both containing a 4~u.c.~thick-STO layer, gives an energy gain 
of about $0.25$~eV per interface u.c. which indicates a 
tendency towards an inhomogeneous spatial distribution of oxygen vacancies in LAO/STO.

It is expected 
that areas with an increased density of oxygen vacancies allow for the formation of ferromagnetic puddles,
as was recently observed by Bert~{\it et.al.}~\cite{bert}. Their 
respective collective magnetic moments are a likely candidate for the source of the superparamagnetic behavior 
observed by Li {\it et al.}~\cite{li}. As compared to the ferromagnetically ordered structure, 
the total energy of the ($2 \times 1$)-configuration with antiferromagnetically
ordered interface increases by 0.36~eV/u.c. which implies a high stability of the ferromagnetic 
state.

\begin{table}[b]
\caption{\label{tab1} Calculated magnetic moments of interface Ti ions nearest to the O-vacancy,
the exchange splitting ($\Delta$) and the parameter $I\rho(\varepsilon_F)$
of the Stoner criterion for (LaAlO$_3$)$_4$/(SrTiO$_3$)$_n$ heterostructures with different $n$.
\\}

\begin{ruledtabular}
\begin{tabular}{llllllll}
n  & $m_{\rm Ti}$($\mu_B$) & $\Delta$(eV)&  $I\rho(\varepsilon_F)$ \\
\hline
1.5 & 0.31 & 1.15 & 3.89 \\
2.5 & 0.33 & 0.82 & 1.26 \\
3.5 & 0.34 & 0.98 & 1.64 
\end{tabular}
\end{ruledtabular}
\end{table}

The ferromagnetic ordering can be examined in the framework of the Stoner model for ferromagnetism.
This model treats the stabilization of the ferromagnetic state as a result of the difference between the 
reduction of the Coulomb interaction for the $3d$ electrons with parallel spins of neighbouring Ti ions and the increase 
of the kinetic energy caused by the widening of the $3d$-bands for the electrons of the same spin \cite{tsymbal}. 
The condition for the appearance of ferromagnetism is based on the Stoner criterion $I\rho(\varepsilon_F)>1$.
The interaction $I=\Delta/m$ parameterizes the exchange splitting $\Delta$,
and $m$ is the magnetic moment of Ti in the ferromagnetic state (per 
$\mu_B$); $\rho(\varepsilon_F)$ is the paramagnetic density of states at the Fermi level. The parameters 
appearing in the Stoner criterion are derived from the results of the GGA+$U$-calculations and presented in 
Table~\ref{tab1}. The Stoner condition is satisfied irrespective of the thickness of the STO-layer, which 
implies that the ferromagnatic state is favourable in these systems although the magnetic moment and exchange 
splitting slightly decrease with increasing STO thickness, consistent with Ref.~\onlinecite{tsymbal}. 
The excess $3d$ charge in the vacancy-doped supercells 
leads to a substantial increase
of $\rho(\varepsilon_F)$ up to $\sim 1.5-1.8$~eV$^{-1}$ as compared to the pure systems with
$\rho(\varepsilon_F)\approx 0.5-0.7$~eV$^{-1}$. The enhanced $\rho(\varepsilon_F)$ contributes 
to the strong increase of the Ti magnetic monents in the oxygen-deficient systems.

\begin{figure}[t]
\epsfxsize=8.0cm {\epsfclipon\epsffile{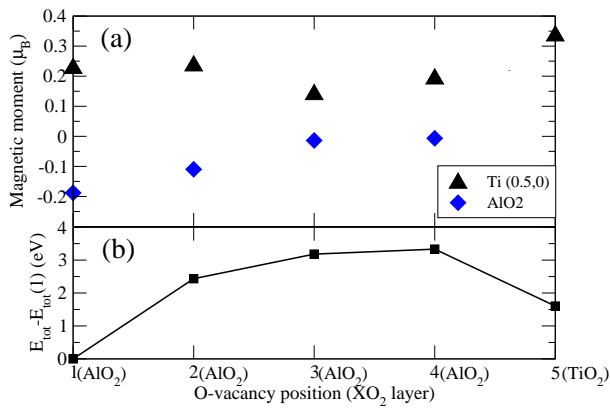}}
\caption{(a) Magnetic moments of the interface Ti and AlO$_2$-layer (per u.c.) in (2STO)/(4LAO)-supercells 
with a change of the position of the O-vacancy along the (001)-direction between the top
AlO$_2$-layer (left) and interface TiO$_2$ (right). (b) Variation of the total energy $E_{\rm tot}-E_{\rm tot}(1)$
with the change of the position of the O-vacancy along the (001)-direction. Here $E_{\rm tot}(1)$ is
the total energy for the system with an O-vacancy located in the top AlO$_2$ surface.
} \label{fig4}
\end{figure}

O-vacancies strongly influence the electronic structure of the Ti $3d$ states: the excess charge
originating from the eliminated O atom in the interfacial TiO$_2$ plane leads to a redistribution in the 
occupancy 
of the five $3d$ orbitals. The contribution of the $3d$ $e_g$ orbitals to the magnetic moment formation is 
rather insignificant. In contrast, a substantial amount of the excess electron 
charge is transferred to the $t_{2g}$ spin-up orbitals (Fig.~\ref{fig2}). 
In a (2$\times$1)-unit cell, the location of the O-vacancy along 
the $y$-direction between the two Ti-atoms (see Fig.~\ref{fig1}(b)) leads to symmetry breaking, 
splitting the two $t_{2g}$ (3d$_{yz}$ and 3d$_{xz}$) orbitals.
Due to the random 
distribution of O-vacancies along the $x$- and $y$-directions in LAO/STO samples, the electron charge is 
assumed to occupy both, the $3d_{yz}$ and the $3d_{xz}$ state---yet the dominant contribution to the 
magnetic moment has to be ascribed to the $3d_{xy}$ spin-up occupancy.

Oxygen vacancies may appear not only at the interfacial TiO$_2$ layer but also on the surface 
and in the bulk of the LAO layer. 
Consistent with the previous calculations ~\cite{zhang} we find that 
the lowest total energy is achieved with the vacancy located in the top
AlO$_2$-surface layer (Fig.~\ref{fig4}(b)), with an energy gain 
of about 1.5~eV as compared to a vacancy in the interface TiO$_2$-layer.
The $25$~\%-concentration of surface vacancies
produces two electrons per (2$\times$1)-unit cell, with one electron transferred to the interface and
another one hybridized at the AlO$_2$ surface close to a central O vacancy, similarly to \cite{elfimov}. 
Our calculations show that
the placement of an O-vacancy in the AlO$_2$-planes of the LAO-layer still induces a significant magnetic
moment on the interface Ti atoms (Fig.~\ref{fig4}(a)). 
This magnetic moment originates from the spin polarization of the occupied Ti-$3d$ states due to the 
electron charge transferred from the O-vacancy site. 
Interestingly, when the O-vacancy lies close to the surface of LAO we find a sizable magnetic moment 
in the AlO$_2$-layer
with the O-vacancy (Fig.~\ref{fig4}(a)), similarly to that obtained when the O-vacancy is 
placed in the interfacial TiO$_2$ layer. 
This magnetic moment is due to the exchange splitting of the not fully occupied local 
oxygen-vacancy-centered state of a mixed Al $s$--$p$-character which produces quasi-one-dimensional
electronic bands and ferromagnetic moments 
in the AlO$_2$-layer antiparallel to that of the interface Ti atom~\cite{paper2}. 

Recent studies of the LAO/STO interfaces distinguish two different types of charge carriers in terms of 
the interfacial localized Ti $3d$ electrons and the mobile $3d_{xy}$ electrons of several distant Ti 
layers~\cite{popovic}, or they relate magnetism to the interfacial electrons produced by the electronic 
reconstruction  and associate superconductivity to the 
charge carriers induced by O-vacancies~\cite{dikin}. Our findings open a new perspective: both the magnetism 
and the superconductivity are due to the interfacial Ti $3d$ electrons. The magnetism, however, is a result of the 
spin splitting of the populated electronic states induced by
O-vacancies, while the metallic behavior of the interface results from the 2D 
electron liquid caused by the electronic reconstruction. The metallic state  has been related to a 
superconducting state below 
300~mK and the predicted scenario suggests that the corresponding charge carriers move in regions of small or 
vanishing O-vacancy concentrations. 

{\it Note added during revision:} After the submission of the manuscript we were informed that Bert~{\it et.~al.}~\cite{bert}
in fact observed submicrometer puddles of ferromagnetism in the presence of the superconducting state at the LaAlO$_3$/SrTiO$_3$
interface. An alternative scenario has recently been proposed by K.~Michaeli~{\it et.~al.}~\cite{michaeli} who suggest
a homogeneously polarized interface layer which is magnetically coupled to a second superconducting layer.  

The authors acknowledge helpful discussions with R.~Ashoori, I.S.~Elfimov, S.S.~Jaswal, Lu Li, L.~Klein, I.V.~Stasyuk, and S.~Seri. 
This work was  supported by the DFG (TRR~80), the EC (oxIDes), Nebraska MRSEC (NSF DMR-0820521),
NSF-EPSCoR (EPS-1010674), the A.~von~Humboldt Foundation 
and the Ministry of Education and Science of Ukraine (Grant No.~0110U001091). 
Financial support from the CFI, NSERC, CRC, and the Max Planck-UBC Centre for 
Quantum Materials are gratefully acknowledged. Grants of computer time from the 
Leibniz-Rechenzentrum M\"unchen through the HLRB project h1181 are thankfully acknowledged.


\begin{references}
\bibitem{paper2} The details of the spin and charge distribution will
be published separately.
\end{references}
\end{document}